\documentclass[12pt]{article}
\usepackage{graphicx}
\usepackage{epsfig}
\usepackage{amsmath}
\usepackage{amssymb}
\usepackage{setspace}

\def\bfi{\begin{figure}}
\def\efi{\end{figure}}

\def\bea{\begin{eqnarray}}
\def\eea{\end{eqnarray}}
\def\beq{\begin{eqnarray}}
\def\eeq{\end{eqnarray}}
\def\bq{\begin{eqnarray*}}
\def\eq{\end{eqnarray*}}
\def\be{\begin{equation}}
\def\ee{\end{equation}}
\def\bm{\begin{math}}
\def\me{\end{math}}

\def\prb#1#2#3{Phys. Rev. B \textbf{#1}, #2 (#3)}

\begin{document}

\begin{center}
{\Large \bf Scaling Behavior of Response Functions in the Coarsening Dynamics of Disordered Ferromagnets}
\vskip1cm
E. Lippiello$^1$, A. Mukherjee$^2$, Sanjay Puri$^2$ and M. Zannetti$^3$ \\
\vskip0.5cm
$^1$Dipartimento di Scienze Ambientali, Seconda Universit\'a di Napoli,
Via Vivaldi, Caserta, Italy. \\
$^2$School of Physical Sciences, Jawaharlal Nehru University,
New Delhi--110067, India. \\
$^3$Dipartimento di Matematica e Informatica and CNISM, Unit\'a di Salerno,
Universit\`a di Salerno, via Ponte don Melillo, 84084 Fisciano (SA), Italy. \\
\end{center}

\begin{abstract}
We study coarsening dynamics in the ferromagnetic random bond Ising model in $d=1,2$. We
focus on the validity of super-universality and the scaling properties of the response
functions. In the $d=1$ case, we obtain a complete understanding of the evolution, from
pre-asymptotic to asymptotic behavior. The corresponding response function shows
a clear violation of super-universality. Further, our results for $d=1,2$ settle the
controversy regarding the decay exponent which characterizes the response function.
\end{abstract}

\newpage

A system undergoes phase ordering when it is suddenly quenched from a high temperature
to below the critical point $\left(T < T_c\right)$ at time $t=0$. The basic feature of
this process is relaxation
via domain coarsening~\cite{Bray,Puri}. In a system without disorder, this is a dissipative
and scale-free phenomenon. The only relevant length scale is the typical domain size, 
which grows as a power law $L(t) \sim t^{1/z}$, where $z$ is the dynamical exponent. The evolving system is
characterized by spatio-temporal correlation functions of the order parameter, which exhibit scaling 
properties. For example, the generic two-time observable ${\cal O}(t,s)$ is expected to scale as
${\cal O}(t,s) = L^{\alpha}(s)f_{{\cal O}}\left[L(t)/L(s)\right]$,
where $(t,s)$ are a pair of times after the quench. We would like to obtain a good understanding of
exponents like  $\alpha$, and the scaling functions $f_{{\cal O}}(x)$~\cite{Bray,Puri}.

There has also been much interest in the ordering of systems with quenched
disorder but without frustration, where the pattern of ergodicity and symmetry breaking
below the critical point is the same as that of the pure system.
This includes ferromagnets subjected to random external fields or
with random exchange couplings, modeled by the \emph{random field Ising model} (RFIM) or
the \emph{random bond Ising model} (RBIM) \cite{sp04}. 
The primary effect of disorder is to create energy barriers slowing down the coarsening process. The 
dependence of these barriers on $L$ determines the nature of the asymptotic growth law, which is 
usually logarithmic or power-law with a disorder-dependent exponent.

Apart from the growth law, a key question is how disorder affects the scaling function
$f_{\cal O}(x)$, namely whether it enters only through the growth law or if there is also an explicit dependence. In the first case, $f_{\cal O}(x)$ is the same for the pure and disordered systems -- this is referred to as {\it super-universality} (SU)~\cite{su}. In the second case, the scaling relation
generalizes to
\be
{\cal O}(t,s) = L^{\alpha}(s)f_{{\cal O}}\left[L(t)/L(s),L(s)/L^*\right],
\label{2}
\ee
where $L^*$ is a disorder-dependent scale. Typically, a scaling function of the above form implies
a crossover from pre-asymptotic $[L(s) \ll L^*]$ to asymptotic $[L(s) \gg L^*]$ behavior.

The validity of the SU hypothesis has been mainly investigated for the
equal-time correlation function or structure factor, and it is found to hold in a wide
variety of cases~\cite{pcp91,ppr04}. Recently, the investigation of SU has been extended
by Henkel and Pleimling (HP) \cite{HP1,HP2} to the autocorrelation and autoresponse functions
in the $d=2$ RBIM, again confirming its validity. So far, no crossover of the type implied
by Eq.~(\ref{2}) has been reported, except for the $d=1$ RFIM~\cite{decandia}.

In this letter, we present a comprehensive study of the response functions
in the ordering dynamics of the $d=1$ and $d=2$ RBIM. In the $d=1$ case, we obtain
a complete theoretical picture of the early- and late-stage dynamics. We
demonstrate the existence of a crossover in the scaling functions, and the consequent violation
of SU. Furthermore, from our study of the zero-field-cooled susceptibility $\chi (t,s)$,
we find that the corresponding decay
exponent $a=0$, as in the pure system. In the $d=2$ case, it is more difficult to make a
clear statement on the issue of SU. However, the data for $\chi (t,s)$
conclusively shows that the scaling exponent $a$ depends on dimensionality $d$, and is consistent with
the phenomenological formula $a = (d-1)/(2z)$. In fact, the presence of disorder enables us
to fix this relation with higher precision than has been previously possible in the pure case.

Let us first present results for the $d=1$ RBIM, which is defined by the
Hamiltonian ${\cal H} = -\sum_{i=1}^{N} J_{i}\sigma_i\sigma_{i+1}$, where the
spins $\sigma_i=\pm 1$. The ferromagnetic couplings
$J_{i}$ are independent random variables, uniformly distributed in the interval
$(1-\epsilon,1+\epsilon)$ with $0 \leq \epsilon \leq 1$. This system undergoes ordering only
at $T=0$. However, the $T=0$ quench must be avoided to prevent the
dynamics from getting stuck in metastable states. The way out is to quench to a
temperature high enough to drive activated coarsening, but low enough
to inhibit the nucleation of equilibrium fluctuations. Then, as long as $L(t)$ is smaller
than the equilibrium correlation length $\xi (T)$, one observes the same coarsening behavior
as in the $T=0$ quench.

\begin{figure}[ht]
\begin{center}
\includegraphics*[width=0.5\textwidth]{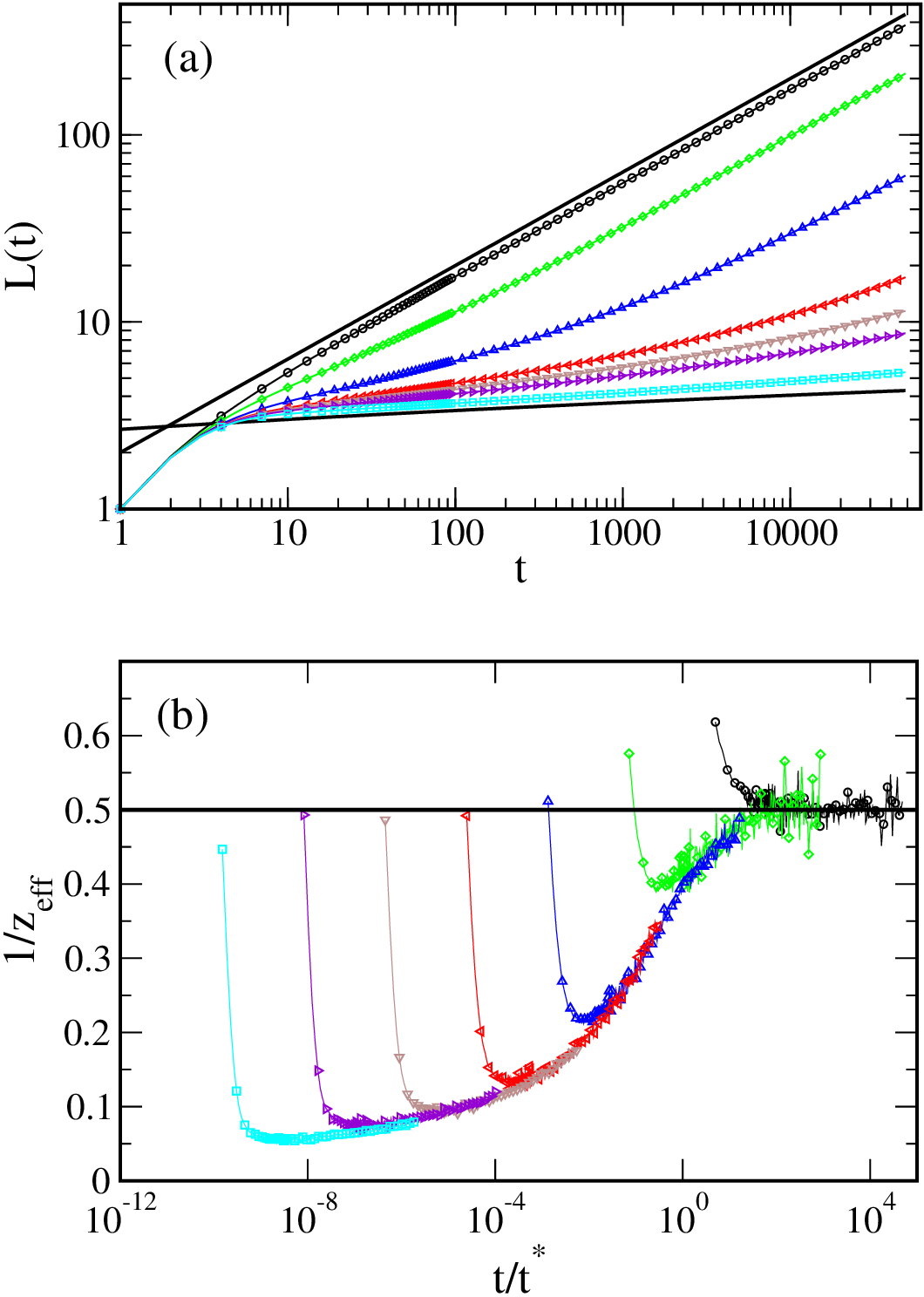}
\end{center}
\caption{\label{fig0} (a) Domain growth law for the $d=1$ RBIM. The disorder amplitude
$\epsilon$ ranges from $0$ to $0.3$ in steps of $0.05$ (top to bottom). The straight
lines depict the laws $L(t) \sim t^{1/2}$; and $L(t) \sim L(t_0) (t/t_0)^{1/z(\epsilon/T)}$
with $\epsilon/T=3$, $t_0=11$ and $1/z(\epsilon/T)=0.05$ (see text). (b) Plot of $1/z_{\rm eff}$
vs. $t/t^*$ for the data sets in (a).}
\end{figure}

We have performed simulations of the Glauber-RBIM with different values of $\epsilon$
on a system of linear size $N=10^5$, up to $t_{\rm{max}}=50000$ Monte Carlo
steps (MCS). At $t=0$, the system is quenched from $T=\infty$ to $T= 0.05$. For each value of
$\epsilon$, averages have been taken over $2000$ independent realizations of disorder and initial conditions. We have studied the growth law [Fig.~\ref{fig0}(a)], measuring $L(t)$ as the inverse
density of defects. This is obtained by dividing the number of sites with at least one
oppositely-aligned neighbor by the total number of sites~\cite{f1}.

Figure~\ref{fig0}(a) shows the existence of three regimes. At {\it early} times, the growth is
very rapid and independent of $\epsilon$. This is the time regime where all micro-domains seeded
by the random initial condition are eliminated with a few flips. Then, the interfaces get trapped in
the local energy minima, namely on the weakest nearby bond. This is followed by the {\it intermediate} regime, where growth is slowed down by activated escape over the barriers. This regime is absent in 
the pure case and lasts up to the time needed to overcome the largest barrier $4\epsilon$, $t^* \sim \exp \left(4\epsilon/T\right)$. Finally, in the {\it asymptotic} regime, interfaces are effectively free and $L(t)$ grows algebraically with $z=2$, as in the pure case. This is clearly shown by the plot of the effective exponent $1/z_{\rm eff} = d\left(\ln L\right)/d \left(\ln t\right)$  vs. $t/t^*$ in Fig.~\ref{fig0}(b). Here, the early-time behavior corresponds to the initial fast drop, and the
intermediate regime corresponds to the ensuing climb of the curves. In the asymptotic regime, 
the effective exponent becomes constant at $z_{\rm eff}=2$ for $t/t^* \geq 1$.

Denoting the duration of the early regime by $t_0$, and integrating $1/z_{\rm eff}$ from $t_0$ 
onward (where it depends only on $t/t^*$), the scaling form of the growth law
$L(t)= L(t_0){\cal L}(t/t^*,t_0/t^*)$ is found, with 
\be
{\cal L}(t/t^*,t_0/t^*) = \exp \left[ \int_{t_0/t^*}^{t/t^*} dx \, {1 \over x z_{\rm eff}(x)}\right] .
\label{3}
\ee
Hence, for values of $\epsilon$ such that the intermediate regime is long enough to allow
for both $L(t) \ll L(t^*)$ and $1/z_{\rm eff}(x)$ small and slowly varying, one has a crossover
from a power-law behavior (with a disorder-dependent exponent) to the power law of the pure case:
\begin{align}
\label{4bis}
  L(t) = \begin{cases} L(t_0)(t/t_0)^{1/z(\epsilon/T)},   &\text{for}~~ t_0 < t \ll t^* , \\ 
D(\epsilon/T) (t/t^*)^{1/2}  ,   &\text{for}~~ t \gg t^* . \end{cases} 
\end{align}
Here, $1/z(\epsilon/T)$ is the minimum of $1/z_{\rm eff}(x)$ for a given value of $\epsilon/T$,
and $D(\epsilon/T)=L(t_0)[\exp \int_{t_0/t^*}^{1}dx (xz_{\rm eff})^{-1}]$ is the disorder-dependent
diffusion constant. The two limiting power-law behaviors are denoted by solid lines in
Fig.~\ref{fig0}(a).

\begin{figure}[ht]
\begin{center}
\includegraphics*[width=0.5\textwidth]{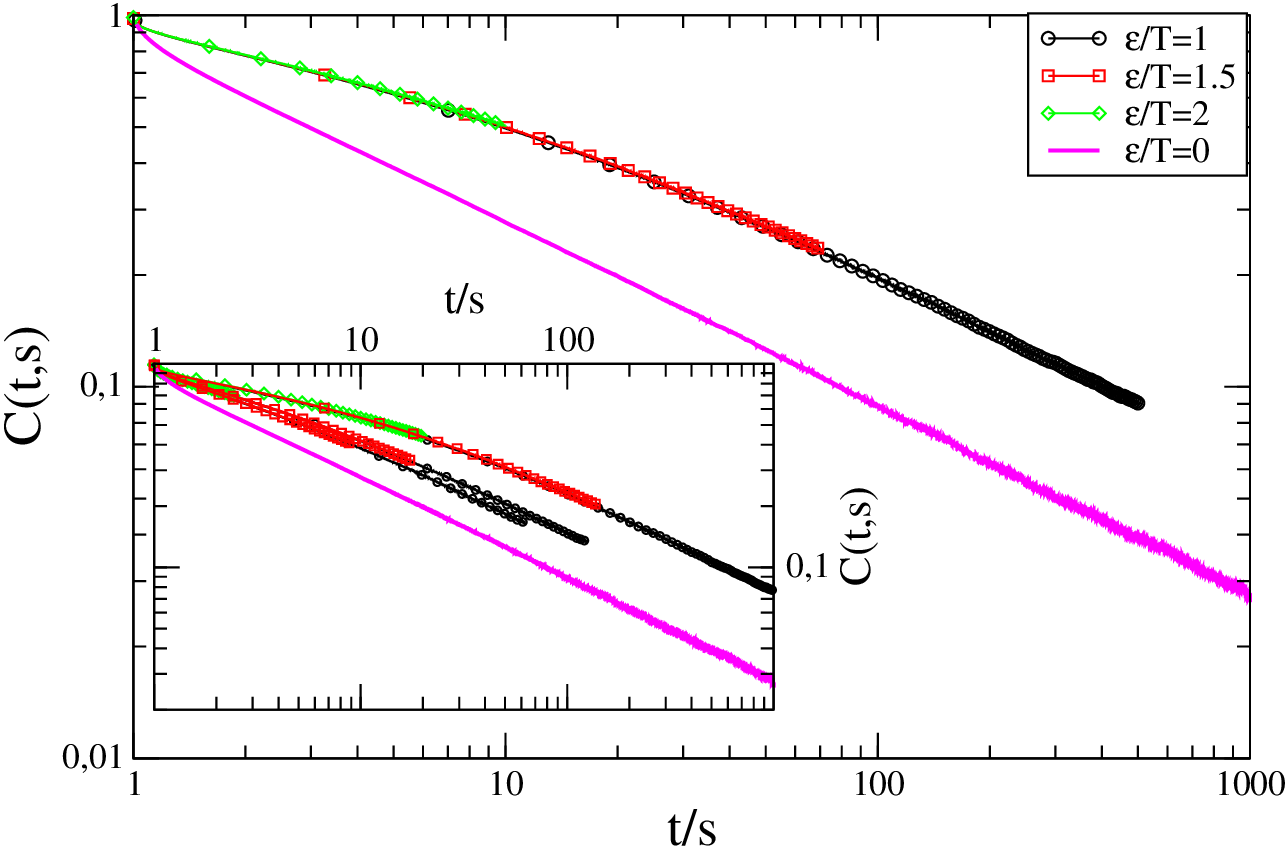}    
\end{center}
\caption{\label{fig1} Main panel: The upper curve plots $C(t,s)$ vs. $t/s$ with
$q=t^*/s=1.1$ and $s = 50,200,800$. The lower curve corresponds to the pure case.
Inset: The same plot with three values of $q=1.1,0.27,0.07$ (from top to bottom).}                 \end{figure}

Our next step is to check for the validity of SU in the $d=1$ RBIM.
We have calculated the autocorrelation and autoresponse functions, which show scaling behavior 
as in Eq.~(\ref{2}) with $L^*=L(t^*)$. In Fig.~\ref{fig1}, we plot $C(t,s) = \langle
\sigma_i(t)\sigma_i(s) \rangle$ vs. $t/s$ for different values of $\epsilon$ and $s$, and a fixed value of the ratio $q=t^*/s = 1.1$. This plot shows an excellent data collapse. According to Eq.~(\ref{2}), the collapse should occur in the plot of $C(t,s)$ vs. $L(t)/L(s)$, but it is straightforward to show from Eq.~(\ref{3}) that $L(t)/L(s)$ and $L(s)/L^*$ depend on the time arguments through the ratios $t/s$ and $s/t^*$. In the inset, we plot $C(t,s)$ vs. $t/s$ for different values of $q$. In this case, there are different scaling curves for each $q$, demonstrating that the scaling function violates SU. 
Further, the curves tend towards that for the pure case as $q \rightarrow 0$.

As mentioned earlier, HP \cite{HP1,HP2}, have studied the linear (integrated)
response function for the $d=2$ RBIM. The study of two-time quantities provides a novel testing ground for aging and SU. The response function determines the effect on the local magnetization at the time $t$, due to a small, constant and site-dependent external field switched on during a time interval preceding $t$. Different response functions
correspond to different choices of the time interval~\cite{CLZ}. 
For instance, HP study the {\it thermoremanent magnetization} (TRM), which 
corresponds to switching the field on during $(0,s)$ with $s <t$. Here, we study 
the {\it zero-field-cooled susceptibility} $\chi(t,s)$, corresponding to the field acting in
the interval $(s,t)$. The scaling behavior of this quantity is an unsettled issue 
even for pure systems. We know that the scaling form $\chi(t,s) \sim s^{-a}f_{\chi}[L(t)/L(s)]$
is obeyed in the aging regime, but the exponent $a$ is still a matter of controversy~\cite{CLZ,zannetti}. 
Our results in this paper settle this issue for both pure systems and the RBIM.
  
There are two different arguments for $a$, which originate in the context of pure systems.
The first picture is based on the idea that the response comes entirely from the paramagnetic
spins at the interfaces~\cite{Barrat}. Therefore, the response function
(per spin) should decrease at the same rate as the interface density $L^{-1}(t)$,
implying
\be
az=1 ,
\label{az0}
\ee
independent of $d$~\cite{HPGL}. Numerical support for Eq.~(\ref{az0}) has been obtained by measuring
the TRM \cite{HPGL,HP0}.

In the second picture, there is another mechanism in addition to the paramagnetic response, whereby
domains as a whole grow so as to minimize the magnetic energy.
This picture, which is supported by some analytical~\cite{anal1,anal2} and numerical results~\cite{num,castellano} 
for $\chi (t,s)$, yields the phenomenological formula:
\begin{align}
\label{4}
az = \begin{cases} (d-1)/2,   &\text{for}~~ d < 3,  \\ 1,   &\text{for}~~ 
                     d \geq 3. \end{cases} 
\end{align}
The different behaviors for $d <3$ and $d > 3$ are related to the roughening of the
interfaces~\cite{castellano}.

Despite considerable numerical effort, it has not been possible to 
clearly decide in favor of Eq.~(\ref{az0}) or Eq.~(\ref{4}) for
pure systems. Considering that there is a large discrepancy between the two expressions for $d=1,2$, 
one may wonder why this is so. 
The root of the problem lies in the use of different response functions. We have 
argued~\cite{CLZ} that the TRM is affected by an extended crossover, 
which prevents the observation of the asymptotic scaling behavior in the simulations \cite{HPGL}.
On the other hand, the zero-field-cooled susceptibility is free of this shortcoming \cite{CLZ}.

\begin{figure}[ht]
\begin{center}
\includegraphics*[width=.5\textwidth]{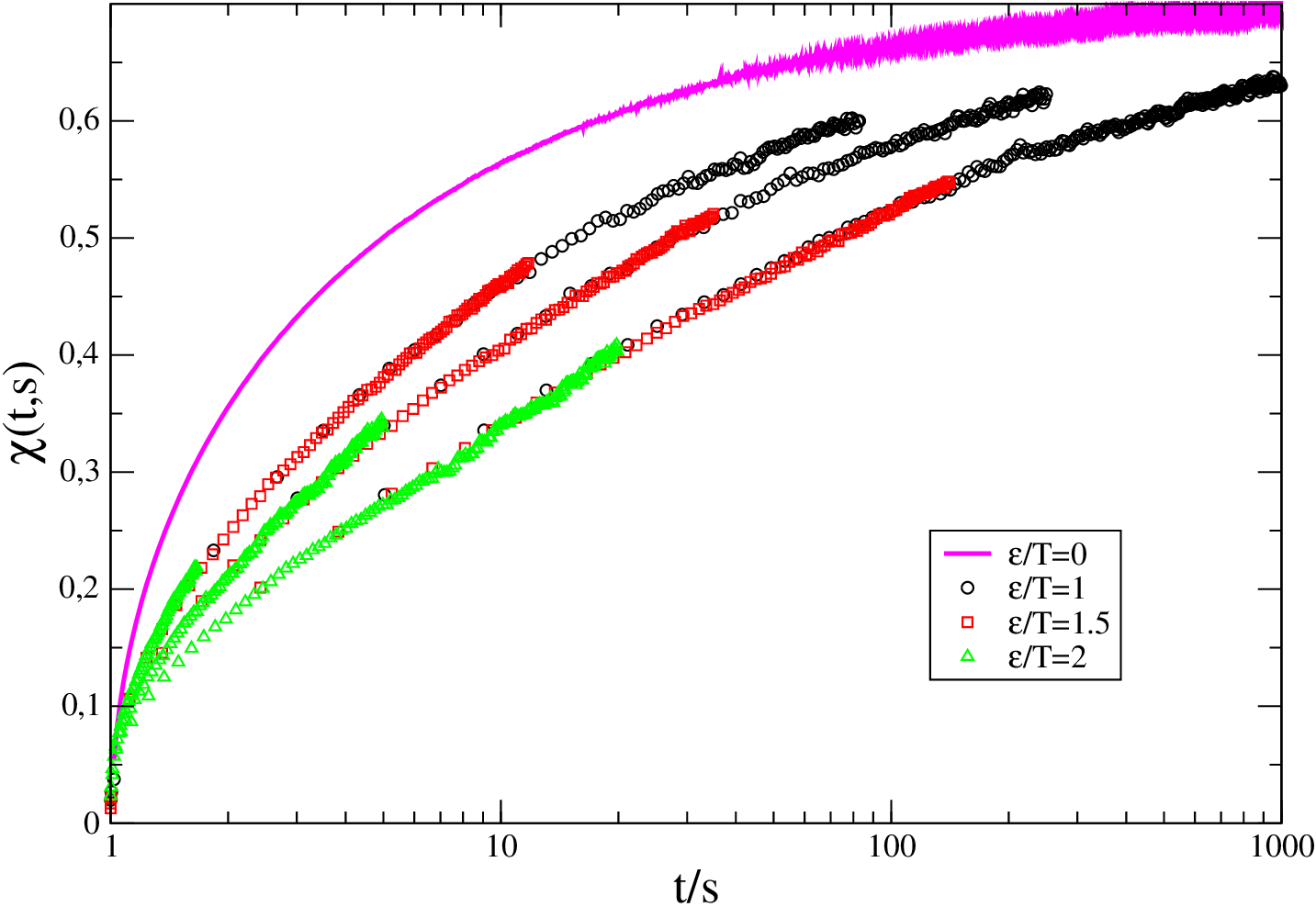}    
\end{center}
\caption{\label{figchi1} Plot of $\chi(t,s)$ vs. $t/s$ for the $d=1$ RBIM with $q=t^*/s=1.1,0.27,0.07$ (bottom to top). We present data for several values of $s$, ranging from 50 to 30000.}
\end{figure}

In the light of the above discussion, let us obtain the exponent $a$ for the $d=1$ RBIM. 
The plot of $\chi(t,s)$ vs. $t/s$ in Fig.~\ref{figchi1} displays the same scaling behavior 
and violation of SU as $C(t,s)$ in Fig.~\ref{fig1}.
Again, the data sets show excellent collapse when $q=t^*/s$ is kept constant. However,
the scaling function depends on $q$ and approaches the pure result as $q\rightarrow 0$. The collapse for
many different values of $s$ in Figure~\ref{figchi1} shows that $a=0$ for the $d=1$ RBIM, consistent with Eq.~(\ref{4}), and as demonstrated analytically in the pure system~\cite{anal1}.

Let us consider next the $d=2$ RBIM. First, we focus on the domain growth law.
Huse and Henley~\cite{HH} have argued that, in this case, the barrier heights scale as 
$\Delta E \sim \Upsilon(\epsilon) L^{\psi}$. The matching of the barrier size with
the thermal energy $\Delta E = T$ introduces a new length scale
into the problem, $L^* = (T/\Upsilon)^{1/\psi}$.
Then, there should be a pre-asymptotic regime for $L(t) \ll L^*$,
with the algebraic behavior $L(t) \sim t^{1/z}$
of the pure system. This is followed by an asymptotic regime for $L(t) \gg L^*$, where 
growth becomes logarithmic $L(t) \sim [\ln (t/t^*)]^{1/\psi}$.
However, it remains controversial whether the crossover is from algebraic to
logarithmic, or from algebraic to algebraic with $z$ dependent
on disorder. A number of experiments on random magnets \cite{exp} have suggested the latter.
Further, Paul et al. \cite{ppr04} have obtained comprehensive MC data for the $d=2$ RBIM, which supports
the latter scenario, at least on numerically accessible length-scales and time scales. However,
Cugliandolo et al.~\cite{Cugliandolo,Iguain} have argued that this is an intermediate regime.

\begin{figure}[ht]
\begin{center}
\includegraphics*[width=0.5\textwidth]{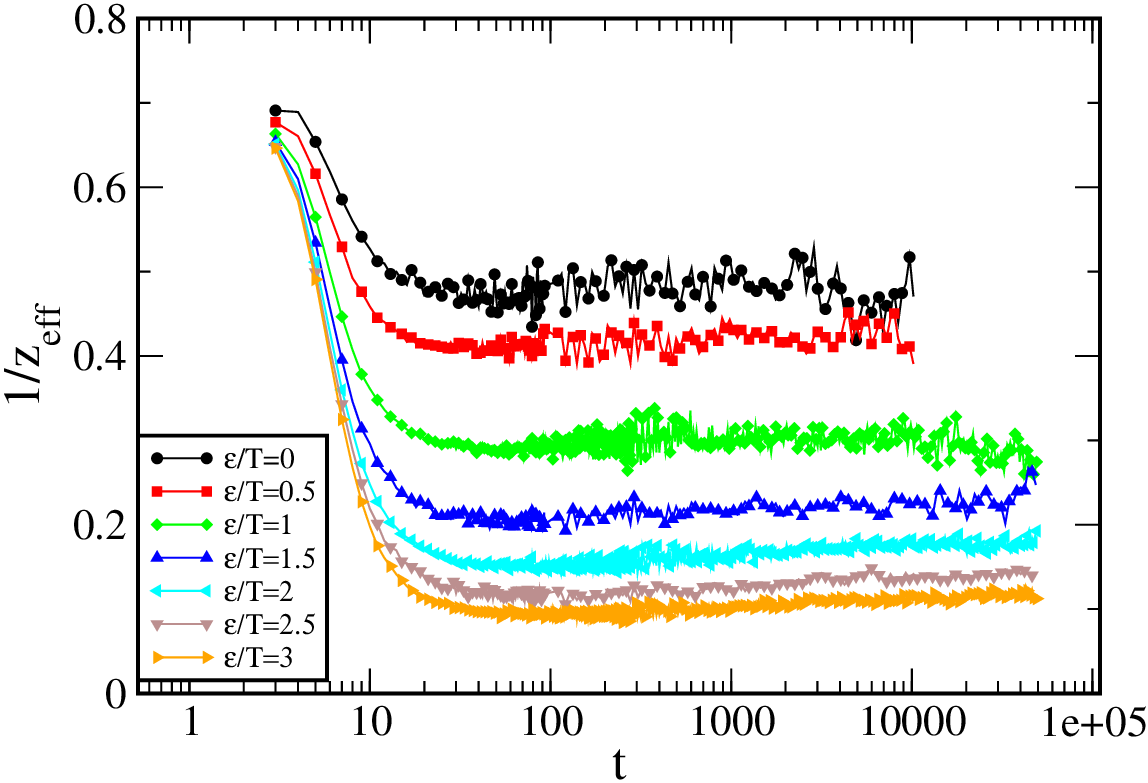}
\end{center}
\caption{\label{zeff_d2} Plot of $1/z_{\rm eff}$ vs. $t$ for the $d=2$ RBIM.}
\end{figure}

We have measured $L(t)$ from the density of defects, as in the $d=1$ case, 
after quenching a $1000^2$ system from $T=\infty$ to $T=0.1$. We average over different disorder
realizations (from $10$ to $30$) for each value of $\epsilon/T$. In Fig.~\ref{zeff_d2}, 
we plot the effective exponent $1/z_{\rm eff}$ vs. $t$, similar to Fig.~\ref{fig0}(b). 
The data sets show the existence of an early regime, between $0$ and $t_0$, followed by a second regime. As in the $d=1$ case, $t_0$ is independent of $\epsilon$, but somewhat larger. For $t > t_0$, $1/z_{\rm eff}$ becomes approximately constant over $3$ decades of time. 
On the time-scales of our simulation, there is no sign of a crossover to 
a logarithmic growth, consistent with earlier studies \cite{exp,ppr04}. The value of the growth exponent 
is consistent with the $z(\epsilon/T) = 2+c\epsilon/T$ behavior proposed by Paul et al.~\cite{ppr04}.
Without tackling the issue of what is the ``truly asymptotic growth law'' in the $d=2$ RBIM,
we stress that we have at least 3 decades of an algebraic growth law. The corresponding exponent is $\epsilon/T$-dependent and spans the range from $z=2$ to $z\simeq 9$. 

Let us turn next to the response functions in the $d=2$ RBIM. The SU of these functions will
be discussed in a later publication. Here, we focus on the decay exponent $a$. HP~\cite{HP2} have obtained $a$
from the TRM for a range of $z$-values. However, their results are not consistent with either Eq.~(\ref{az0}) or Eq.~(\ref{4}). Concerned with the violation of Eq.~(\ref{az0}), HP have argued that the departure is due to the disorder inducing fractality in the interfaces. 
This would modify  Eq.~(\ref{az0}) to $az=d-d_f$, with $d_f$ being the fractal dimensionality 
of interfaces. In order to test this, we have examined the equal-time correlation 
function, which scales as $C(r,t)=f(r/L)$. 
For fractal interfaces, the short-distance decay (or Porod decay) of this quantity is 
$C(r,t) \simeq 1-b(r/L)^{d-d_f}$ for $r/L\ll 1$. We find  no difference between the short-distance
behavior of $C(r,t)$ with and without disorder. Therefore, disorder does not  result 
in fractal interfaces and Eq.~(\ref{az0}), if correct, should also hold with disorder. We 
conclude that the HP data is unexplained due to the difficulty of accessing the asymptotic regime
with the TRM.

\begin{figure}[ht]
\begin{center}
\includegraphics*[width=0.5\textwidth]{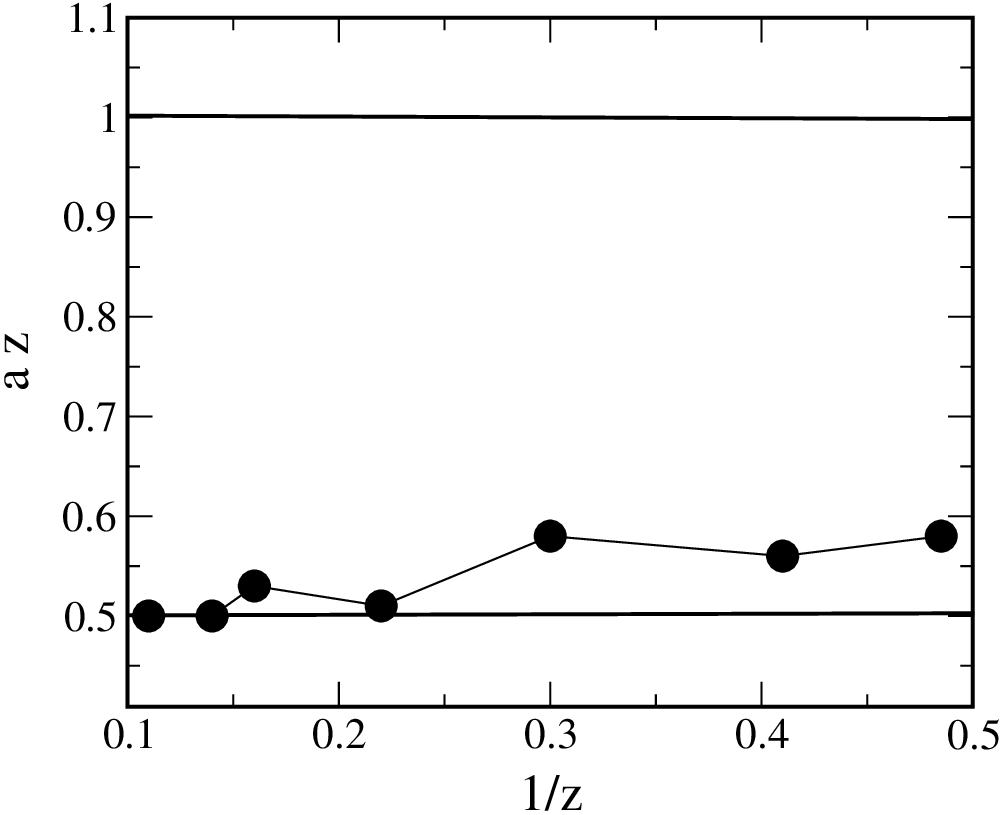}    
\end{center}
\caption{\label{az} Plot of $az$ vs. $1/z$ for the $d=2$ RBIM. The horizontal lines are drawn at
$az=1$ and $az=1/2$.}                                    
\end{figure}

We now present our results for $a$ in the $d=2$ RBIM. For each value of $\epsilon/T$,
we have computed $a$ as the exponent producing the best data collapse of  $L^{az}(s)\chi(t,s)$ vs. $t/s$, for different values of $s$. The measured products $az$ are plotted in Fig.~\ref{az}, together with Eq.~(\ref{az0}) ($az=1$) and Eq.~(\ref{4}) ($az=1/2$). Clearly, our data is consistent with Eq.~(\ref{4}), and completely rules out Eq.~(\ref{az0}).
We stress that both $z$ and $a$ vary with disorder (cf. Fig.~\ref{zeff_d2} for
$z$ vs. $\epsilon$), but the product $az$ seems to be universal.

In summary, we have undertaken a comprehensive study of autocorrelation and response functions in 
the ordering dynamics of the RBIM. In $d=1$, the growth exponent shows a crossover from a
pre-asymptotic disorder-dependent value to the asymptotic value $z=2$, as in the pure case.
The corresponding autocorrelation and response functions violate SU. 
The scaling exponent of $\chi(t,s)$ is $a=0$, consistent with $az=0$ from Eq.~(\ref{4}). 
In $d=2$, after the transient regime, we see an extended regime of power-law growth with $z$ being 
dependent on $\epsilon/T$. More important, we find $az\simeq1/2$, again consistent with Eq.~(\ref{4}). 
In general, the introduction of disorder complicates the domain growth problem. However, it also provides us an excellent opportunity to make a clear assessment of the $az$-relationship. We believe that a further pursuit of phenomenology is of secondary importance at this stage. Now the focus should be on a significant theoretical advance enabling us to ascertain the accuracy of Eq.~(\ref{4}).

\vskip0.25cm
\noindent{\bf Acknowledgments}

MZ acknowledges financial support from PRIN 2007 JHLPEZ. He wishes to thank the Jawaharlal Nehru Institute of Advanced Study and the School of Physical Sciences at the Jawaharlal Nehru University for  hospitality as well as financial support.

\newpage

\begin{thebibliography}{99}
\bibitem{f1} We have checked that this definition is in good agreement with the usual method of measuring $L(t)$ by the half-peak-width of the equal-time structure factor.
\end{thebibliography}
\end{document}